\documentclass[a4paper]{article}

\usepackage[english]{babel}
\usepackage[utf8x]{inputenc}
\usepackage[T1]{fontenc}
\usepackage{layouts}

\usepackage[hidelinks]{hyperref}

\usepackage[a4paper,top=3cm,bottom=2cm,left=3cm,right=3cm,marginparwidth=1.75cm]{geometry}

\usepackage{amsmath}
\usepackage{graphicx}
\usepackage[format=plain,margin=10pt,font=small,labelfont=bf,
labelsep=endash]{caption}

\title{CameraTransform: a Scientific Python Package \\for Perspective Camera Corrections}
\usepackage{authblk}
\author[1]{Richard Gerum}
\author[1]{Sebastian Richter}
\author[1]{Alexander Winterl}
\author[1]{Ben Fabry}
\author[1,2]{Daniel Zitterbart}
\affil[1]{Department of Physics, University of Erlangen-N\"urnberg, Germany}
\affil[2]{Applied Ocean Physics and Engineering, Woods Hole Oceanographic Institution, Woods Hole, USA}

\begin{document}
	\maketitle
	
	\begin{abstract}
		Scientific applications often require an exact reconstruction of object  positions and distances from digital images. Therefore, the images need to be corrected for perspective distortions. We present  \textit{CameraTransform}, a python package that performs a perspective image correction whereby the height, tilt/roll angle and heading of the camera can be automatically obtained from the images if additional information such as GPS coordinates or object sizes are provided. We present examples of images of penguin colonies that are recorded with stationary cameras and from a helicopter.
	\end{abstract}
	
	\section{Introduction}
	Optical recordings such as on-demand images from camera traps, continuous time-lapse images, or video recordings, are a widely used tool in ecology \cite{Gregory2014, Zitterbart2011, 10.2307/3784076}. While such recordings are useful for counting animals and estimating abundances \cite{Lynch2015}, they inherently contain perspective distortions that make it difficult to measure positions and distances. To correct for such distortions and to  map image points to real-world positions, it is paramount to know certain camera parameters. This includes the geographic camera position relative to landmarks in the  scenery, the camera height, tilt/roll angle and heading. These parameters are often difficult or impossible to evaluate in the field at the time of the recording, but they can be reconstructed afterwards if the real-world coordinates of prominent features in the images are known. The mathematical procedure behind this reconstruction is based on simple linear algebra, but the steps to apply the underlying matrix operations to image data can be somewhat involved . 
	
	In this article we present the python package \textit{CameraTransform} that was developed to facilitate post-recording calibration based on single (not stereo) images.  \textit{CameraTransform}  provides various tools to estimate the camera parameters from features present in the image, and transforms point coordinates in the image to  real-world or to geographic coordinates. We explain the mathematical details of  the calibration and transformation,  present calibration examples and provide an analysis of the uncertainty of the procedures.
	
	\section{Camera Matrix}
	All information about the mapping of real-world points to image points are stored in a camera matrix. The camera matrix is expressed in projective coordinates, and can be split into two parts: the intrinsic matrix and the extrinsic matrix \cite{hartley2003multiple}. The intrinsic matrix  depends on the camera sensor and lens, the extrinsic matrix depends on the camera's position and orientation.
	
	\subsection{Projective coordinates}
	Projective coordinates, also known as homogeneous coordinates, are used to represent projective transformations as matrix multiplications \cite{mobius1827barycentrische}. They are a mathematical trick that extends the vector representation of a point with an additional entry. This entry defaults to 1, and all scalar multiples of a vector are considered equal:
	\begin{align}
		\begin{pmatrix}
			x \\
			y \\
			1
		\end{pmatrix}
		\hat{=}  \begin{pmatrix}
			s\cdot x \\
			s\cdot y \\
			s
		\end{pmatrix}
	\end{align}
	For example, the point (5,7) can be represented by the tuple of projective coordinates (5,7,1) or (10,14,2) and so on. The scalar $s$ need not be an integer. 
	Projective coordinates allow us to write the camera projection $\vec{y}$ as:
	\begin{align}
		\begin{pmatrix}
			y_1 \\
			y_2 \\
			1
		\end{pmatrix}
		=   \begin{pmatrix}
			c_{11} & c_{12} & c_{13} & c_{14}\\
			c_{21} & c_{22} & c_{23} & c_{24}\\
			c_{31} & c_{32} & c_{33} & c_{34}
		\end{pmatrix} \cdot
		\begin{pmatrix}
			x_1 \\
			x_2 \\
			x_3\\
			1
		\end{pmatrix}
	\end{align}
	where $\vec{x}$ specifies the point in the 3D world, which is transformed with the camera matrix $C$ to obtain the point in the camera image $\vec{y}$.

	\subsection{Intrinsic parameters}
	To compute the intrinsic matrix entries, we need to know the focal length $f$ of the camera in mm, the sensor dimensions ($w_\mathrm{sensor}\times h_\mathrm{sensor}$) in mm, and the image dimensions ($w_\mathrm{image}\times h_\mathrm{image}$) in pixels. The intrinsic matrix entries are then the effective focal length $f_\mathrm{pix}$ and the centre of the image ($w_{\mathrm{image}}/2$,  $h_{\mathrm{image}}/2$) according to
	\begin{align}
		C_{\mathrm{intr.}} &= 
		\begin{pmatrix}
			f_{\mathrm{pix}} & 0 & w_{\mathrm{image}}/2 & 0 \\
			0 & f_{\mathrm{pix}} & h_{\mathrm{image}}/2 & 0 \\
			0 & 0 & 1 & 0 \\
		\end{pmatrix}\\
		f _{\mathrm{pix}} &= f / w_{\mathrm{sensor}} \cdot w_{\mathrm{image}}
	\end{align}
	Here, the diagonal elements account for the rescaling from pixels in the image to a position in mm on the chip. The off-diagonal elements present an offset, whereby the origin of the image is at the top left corner, and the origin of the chip coordinates is at the centre of the chip.
	
	\subsection{Extrinsic parameters}
	\begin{figure}[b]
		\centering
		\includegraphics{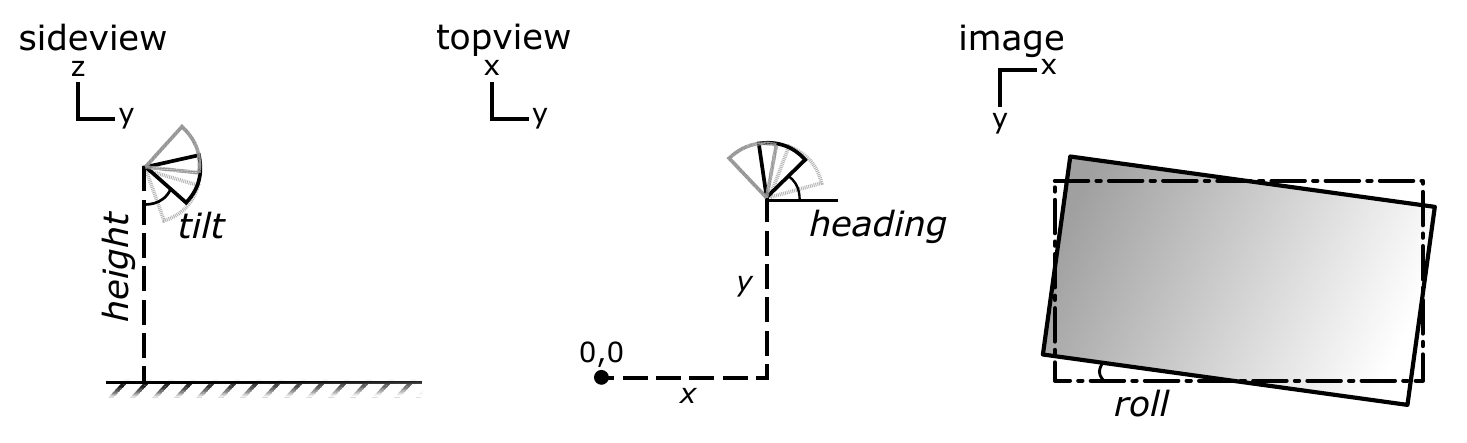}
		\caption{\textbf{Extrinsic camera parameters.}\newline Side view: the height specifies how high the camera is positioned over the ground, the tilt angle specifies how much the camera is tilted against the horizontal. Top view: the offset (x, y) specifies how much the camera is moved from the origin and the heading angle specifies in which direction it is looking. Image: the roll specifies how much the image is rotated around its centre.}
		\label{fig:ExtrinsicParameters}
	\end{figure}
	To compute the extrinsic matrix, we need to know the offset (x,y,z) of the camera relative on an arbitrary fixed real-world reference point (0,0,0) in the three spatial directions. Customarily, the z-coordinate of the reference point is the ground, and z is therefore the height of the camera above ground. Similarly, the x,y plane of our coordinate system is customarily the horizontal plane. We also need to know three angles: the tilt angle $\alpha_\mathrm{tilt}$, which specifies how much the camera is tilted against the horizontal, the heading angle $\alpha_\mathrm{heading}$ which specifies the direction relative to the y-direction in which the camera is heading, and the roll angle $\alpha_\mathrm{roll}$ which specifies how the image is rotated (see Fig.~\ref{fig:ExtrinsicParameters}).
	
	To compute the extrinsic camera matrix, we first need the three rotation matrices and the translation matrix:
	\begin{align}
		R_{\mathrm{tilt}} &= 
		\begin{pmatrix}
			1 & 0 & 0\\
			0 & \cos(\alpha_{\mathrm{tilt}}) & \sin(\alpha_{\mathrm{tilt}}) \\
			0 & -\sin(\alpha_{\mathrm{tilt}}) & \cos(\alpha_{\mathrm{tilt}}) \\
		\end{pmatrix}\\
		R_{\mathrm{roll}} &= 
		\begin{pmatrix}
			\cos(\alpha_{\mathrm{roll}}) & \sin(\alpha_{\mathrm{roll}}) & 0\\
			-\sin(\alpha_{\mathrm{roll}}) & \cos(\alpha_{\mathrm{roll}}) & 0\\
			0 & 0 & 1\\
		\end{pmatrix}\\
		R_{\mathrm{heading}} &= 
		\begin{pmatrix}
			\cos(\alpha_{\mathrm{heading}}) & \sin(\alpha_{\mathrm{heading}}) & 0\\
			-\sin(\alpha_{\mathrm{heading}}) & \cos(\alpha_{\mathrm{heading}}) & 0\\
			0 & 0 & 1\\
		\end{pmatrix}\\
		t &= 
		\begin{pmatrix}
			x\\
			y\\
			-\mathrm{height}
		\end{pmatrix}\\
	\end{align}
	The extrinsic camera matrix then consists of the 3x3 rotation matrix $R$ and the 3x1 translation matrix $t$ side by side, as a 4x4 matrix in projective coordinates.
	\begin{align}
		R &=  R_{\mathrm{roll}} \cdot  R_{\mathrm{tilt}} \cdot  R_{\mathrm{heading}}\\
		T &= R_{\mathrm{tilt}} \cdot  R_{\mathrm{heading}} \cdot t\\
		C_{\mathrm{extr.}} &=  \left(\begin{array}{c|c}
			R & T \\ 
			\hline
			0 & 1
		\end{array}
		\right)
	\end{align}The final camera matrix $C$ is  the product of the intrinsic and the extrinsic camera matrix.
	\begin{align}
		C &=  C_{\mathrm{intr.}} \cdot C_{\mathrm{extr.}}
	\end{align}
	\subsection{Projecting from the World to the Camera}
	Based on the camera matrix $C$ , it is straight forward to see how a real-world point corresponds to a pixel of the acquired image. 
	
	First, the real-world point $\vec{p}_\mathrm{world} (x_1,x_2,x_3)$ is written in projective coordinates:
	\begin{align}
		\tilde{p}_\mathrm{world} = \begin{pmatrix}
			x_1 \\
			x_2 \\
			x_3\\
			1
		\end{pmatrix}
	\end{align}were $\tilde{p}$ denotes the vector $\vec{p}$ in projective coordinates.
	Second, the point $\tilde{p}$ can be projected to the image coordinates:
	\begin{align}
		\tilde{p}_\mathrm{im} = C \cdot \tilde{p}_\mathrm{world}
	\end{align}
	Finally, the point $\tilde{p}_\mathrm{im}$ is converted back from projective coordinates (which has 3 entries) to ``conventional'' coordinates $\vec{p}_\mathrm{im}$ (with two entries) by dividing by the additional scaling factor $s$ (which is the 3rd entry of  $\tilde{p}_\mathrm{im}$):
	\begin{align}
		\vec{p}_\mathrm{im} = \begin{pmatrix}
			\tilde{p}_\mathrm{im_1} / \tilde{p}_\mathrm{im_3} \\
			\tilde{p}_\mathrm{im_2} / \tilde{p}_\mathrm{im_3}
		\end{pmatrix}
	\end{align}
	where the subscript denotes the entry of the vector $\tilde{p}_\mathrm{im}$.
	
	\subsection{Projecting from the Camera back to real-world coordinates}
	While projecting from the 3D real-world to the 2D image is a straight forward matrix multiplication, projecting from the image back to the real-world is more difficult. As the information of the 3rd dimension is lost during the transformation from the real-world to the image, there exists no unique back-transformation. An additional constraint is needed to transform a point back to the 3D world, e.g. one of the 3D coordinates must be fixed. For example: if the real-world point $\vec{p}_\mathrm{world} $ has a fixed  $x_2$ coordinate (for example a mural painting on a vertical wall that is aligned in the y-direction of the coordinate system) and the image coordinates $y_1$ and $y_2$ are given, the back-transformation can be performed as follows:
	\begin{align}
		\begin{pmatrix}
			y_1 \\
			y_2 \\
			1
		\end{pmatrix}
		&=   \begin{pmatrix}
			c_{11} & c_{12} & c_{13} & c_{14}\\
			c_{21} & c_{22} & c_{23} & c_{24}\\
			c_{31} & c_{32} & c_{33} & c_{34}
		\end{pmatrix} \cdot
		\begin{pmatrix}
			s\cdot x_1 \\
			s\cdot x_2 \\
			s\cdot x_3\\
			s
		\end{pmatrix}\\
		& =   \begin{pmatrix}
			c_{11} & c_{12} \cdot x_2 & c_{13} & c_{14}\\
			c_{21} & c_{22} \cdot x_2 & c_{23} & c_{24}\\
			c_{31} & c_{32} \cdot x_2 & c_{33} & c_{34}
		\end{pmatrix} \cdot
		\begin{pmatrix}
			s\cdot x_1 \\
			s \\
			s\cdot x_3\\
			s
		\end{pmatrix}\\
		& =   \begin{pmatrix}
			c_{11} & c_{12} \cdot x_2+c_{14} & c_{13} \\
			c_{21} & c_{22} \cdot x_2+c_{24} & c_{23} \\
			c_{31} & c_{32} \cdot x_2+c_{34} & c_{33} 
		\end{pmatrix} \cdot
		\begin{pmatrix}
			s\cdot x_1 \\
			s \\
			s\cdot x_3
		\end{pmatrix}\\
		&= \tilde{C}  \begin{pmatrix}
			s\cdot x_1 \\
			s \\
			s\cdot x_3
		\end{pmatrix}\\
		\tilde{C}^{-1} \cdot \begin{pmatrix}
			y_1 \\
			y_2 \\
			1
		\end{pmatrix}& =  \begin{pmatrix}
			s\cdot x_1 \\
			s \\
			s\cdot x_3
		\end{pmatrix}
	\end{align}
	This means that the information about the fixed 3D coordinate has to be incorporated in the camera matrix. The inverse of the resulting matrix, when multiplied with the image point in projective coordinates, gives the unknown $x_1$ and $x_3$ entries of the real-world 3D point. After rescaling the vector entries (division by $s$), the known  $x_2$ value is added to the vector to retrieve the real-world coordinates of the 3D point $\vec{p}_\mathrm{world}$. 
	
	The same approach can be used with fixed $x_1$ coordinates or, more relevant for many applications, with fixed  $x_3$  coordinates (i.e. objects on a levelled surface are imaged) (see appendix A).

	\section{Fitting Camera Parameters}
	Often, only the intrinsic camera parameters are known, but not the extrinsic parameters that define the orientation of the camera. The \textit{CameraTransform} package provides several fitting routines that allow users to infer the extrinsic parameters from characteristic features in the image.
	
	In many cases, the heading and position of the camera can be set to 0, as they are only of interest when the camera image needs to be compared to other camera images or when it needs to be cartographically mapped. This leaves only the parameters \textit{height}, \textit{tilt} and \textit{roll} free, unless the camera was properly horizontally aligned, in which case \textit{roll} is zero.
	
	\subsection{Influence of Camera Parameter Uncertainties}
	\begin{figure}[htb]
		\centering
		\includegraphics{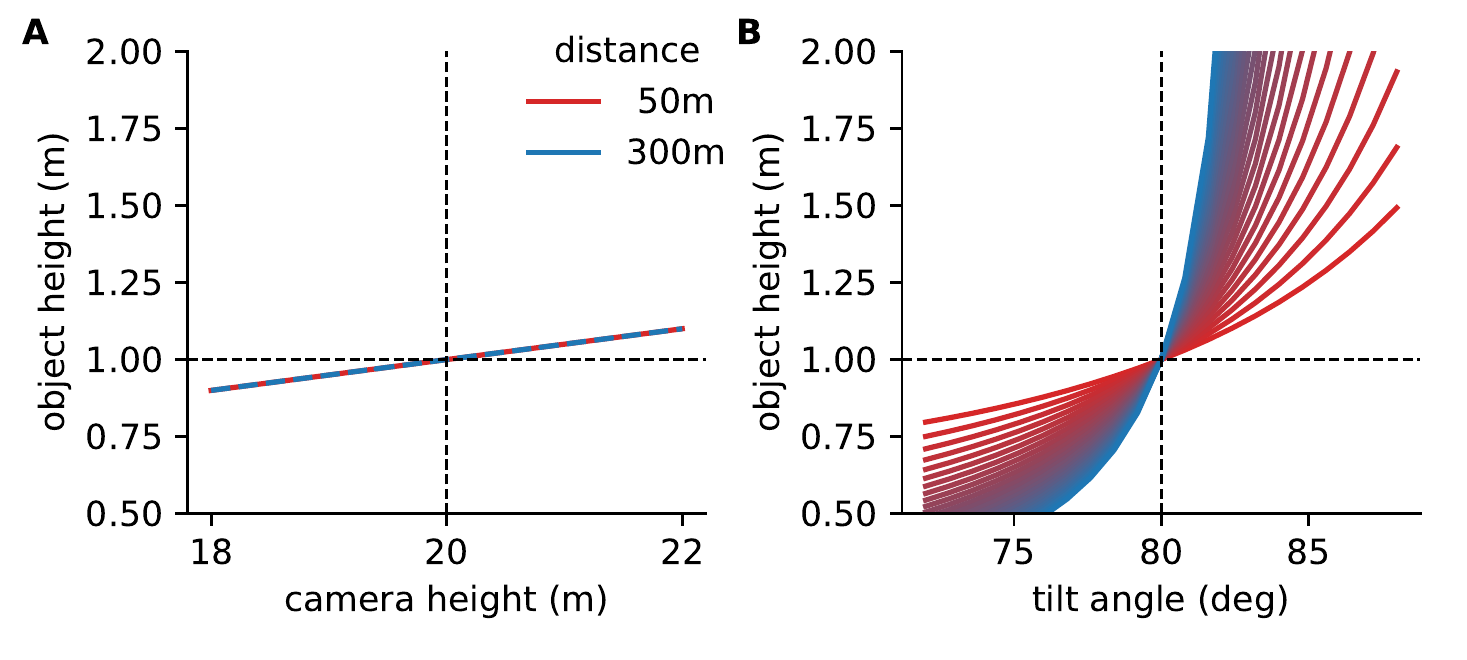}
		\caption{\textbf{Influence of height and tilt angle variation of $\pm$ 10\%.}\newline Objects with a height of 1\,m (dashed line) and different distances (50\,m -- 300\,m) projected to the camera and back to the world with changed camera parameters. A) For variation of the heigh parameter: 20\,m $\pm$ 10\% and B) the tilt parameter: 80° $\pm$ 10\%.}
		\label{fig:variation}
	\end{figure}
	
	To evaluate the sensitivity of the perspective projection with respect to uncertainties in the camera parameters, we computationally place objects of 1\,m height in world coordinates at different distances from the camera (50 -- 300\,m) and project them to the camera image. The positions in the camera image are then projected back to  real-world coordinates using a different parameter set where we vary the camera height and tilt angle. We use a focal length of 14\,mm, a sensor size of 17.3$\times$9.7\,mm with 4608$\times$2592\,px. The camera is placed at a height of 20\,m with a tilt angle of 80°. For the back projection, the height and tilt are varied by $\pm$10\% (Fig.~\ref{fig:variation}). For each parameter configuration, the apparent object height calculated. Since we know the true object height, the reconstructed object height indicates the error that is introduced by uncertainties in the extrinsic camera parameters. 
	
	We find that the apparent object height is robust to variations in camera height regardless of the distance between object and camera (Fig.~\ref{fig:variation}A). By contrast, the apparent object height is sensitive to variations in the camera's tilt angle, especially for objects with larger distance from the camera (Fig.~\ref{fig:variation}B).
	
	\subsection{Fitting extrinsic parameters from object of known height}
	If the true height of objects in the image is know, the camera parameters can be fitted. This works especially well for the tilt angle as it most sensitively affects the apparent object height (Fig.~\ref{fig:variation}B). The input for the fitting routing is a list of base (foot) and top (head) positions of the objects. For fitting, the algorithm projects the foot positions from the image to world coordinates, moves the base positions in z-direction by the known object height, and projects these points back to the camera image. The difference between the input top positions and the back-projected top positions is then minimized with a least-squares fit routine. Optionally, if a horizon is visible  in the image, \textit{CameraTransform} uses the horizon line as an additional constraint for fitting the camera parameters. The error between the user-selected horizon and the fitted horizon is assigned a weight of 50\% of the total error. 
	
	\begin{figure}[htb]
		\centering
		\includegraphics{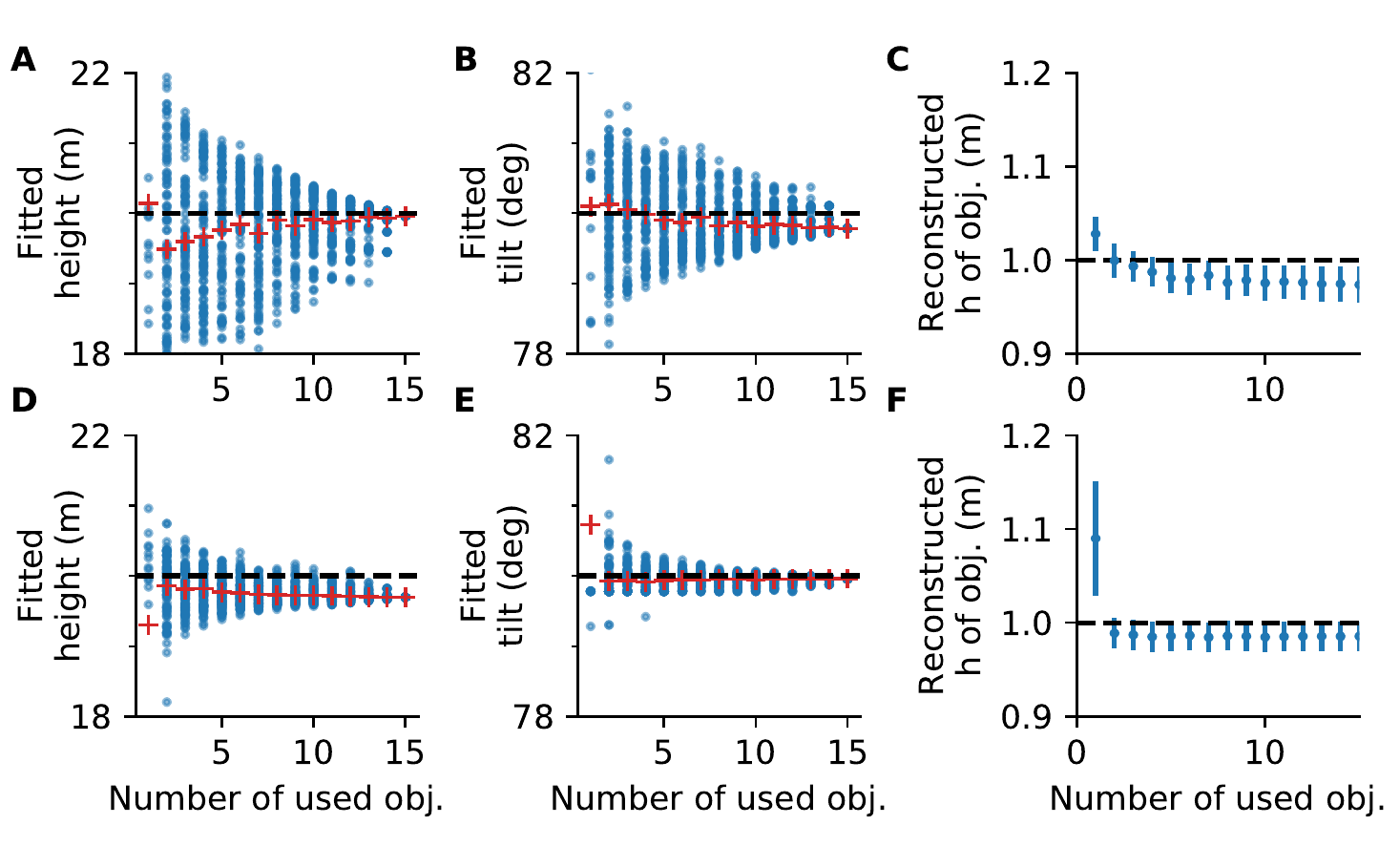}
		\caption{\textbf{Influence of number of object used for fitting.}\newline Top row A-C) without a given horizon, bottom row D-F) with a given horizon. A+D) The fitted camera height and, B+E) the camera tilt angle for different numbers of used objects. For each number of objects a random selection (without replacing) is taken from the clicked objects and the camera matrix is fitted (parameters blue dots). From these the mean is calculated (red crosses). C-F) The error on fitted object heights for different fits (mean$\pm$ std, blue errorbars).}
		\label{fig:ObjectNumber}
	\end{figure}
	To evaluate this method, an artificial image is created using the \textit{CameraTranform} package. We use again a focal length of 14\,mm, a sensor size of 17.3$\times$9.7\,mm with 4608$\times$2592\,px, a camera height of 20\,m and a tilt angle of 80°. 15 rectangles with a width of 30\,cm and a height of 1\,m are placed at distances ranging from 50 to 150\,m. Using the software \textit{ClickPoints}, we mark the base and top positions of these rectangles and provide them as input for the fitting routine. We then investigate how the fitted height and the fitted tilt angle vary with the number of provided objects. We start with only one object and increase the number of objects to 15 . For every iteration, the objects are randomly chosen. The experiment was repeated multiple times with and without a horizon.
	
	The results indicate, as expected, that by including a larger number of objects, the uncertainty of the parameter estimate  (as indicated by the variability between repeated measurements) decreases (Fig.~\ref{fig:ObjectNumber}).  Both, the camera height and the tilt angle can be fitted with considerably less uncertainty if a horizon is provided (Fig.~\ref{fig:ObjectNumber}D,E), compared to parameter estimates without horizon (Fig.~\ref{fig:ObjectNumber}A,B).  The reconstructed object heights ( Fig.~\ref{fig:ObjectNumber}C,F) follow the same pattern and also profit from the horizon information.
	
\begin{figure}[hbt]
		\centering
		\includegraphics{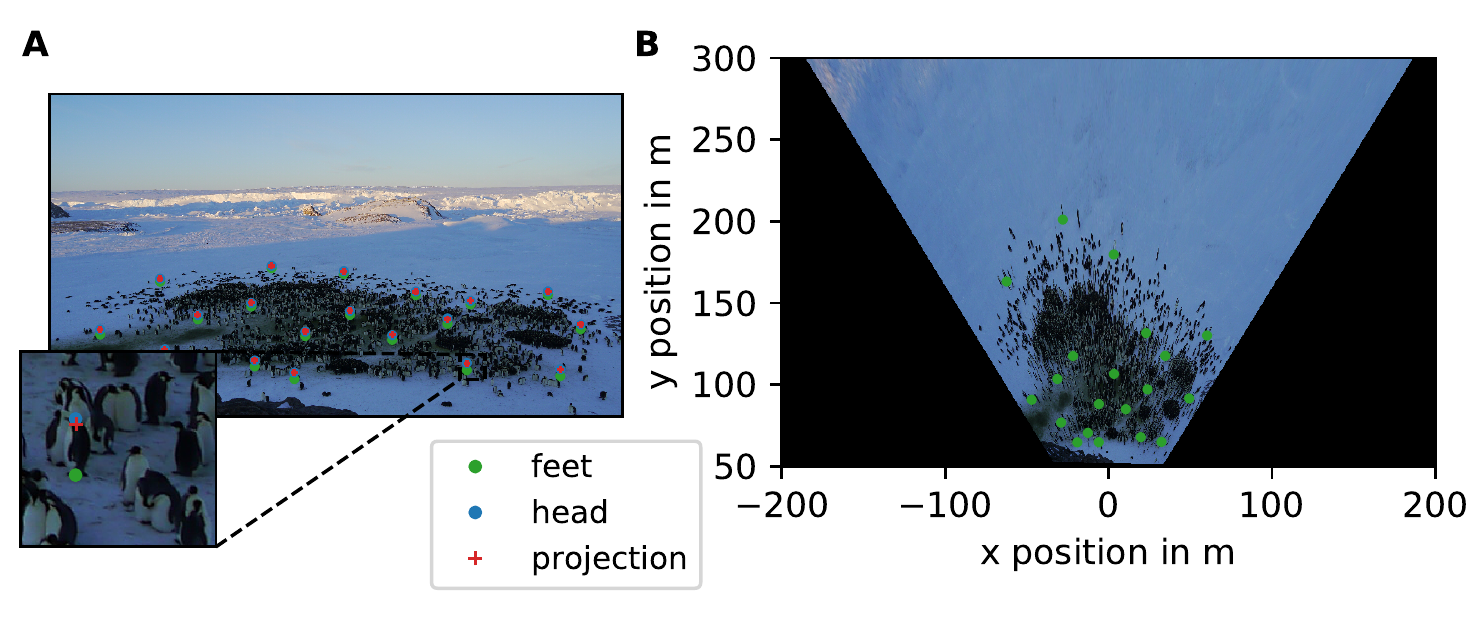}
		\caption{\textbf{Application to real data.}\newline A) Image taken with the MicrObs system of a penguin colony. The feet (green) and head (blue) positions of 20 penguins were manually marked. This data was used to fit the camera perspective (fitted heads: red crosses), which allows to project the image to a top view (B).}
		\label{fig:ObjectNumberRealData}
	\end{figure}

	To demonstrate the fitting procedure, we analyse an image (Fig.~\ref{fig:ObjectNumberRealData}A) from a wide-angel camera overseeing an Emperor penguin colony at Pointe Géologie, Antarctica. The camera was positioned on a nunatak, but no height information was provided. We estimate the extrinsic camera parameters by analysing the feet and head positions of 20 animals, assuming an average penguin height of 1\,m. Fig.~\ref{fig:ObjectNumberRealData}B shows the projected top view after fitting the extrinsic camera parameters. The camera height obtained by the fit is with 23.7\,m close the to the height value of 25.7\,m measured by a differential GPS.
	
	\subsection{Fitting by geo-referencing}
	
	\begin{figure}[htb]
		\centering
		\includegraphics[width=\textwidth]{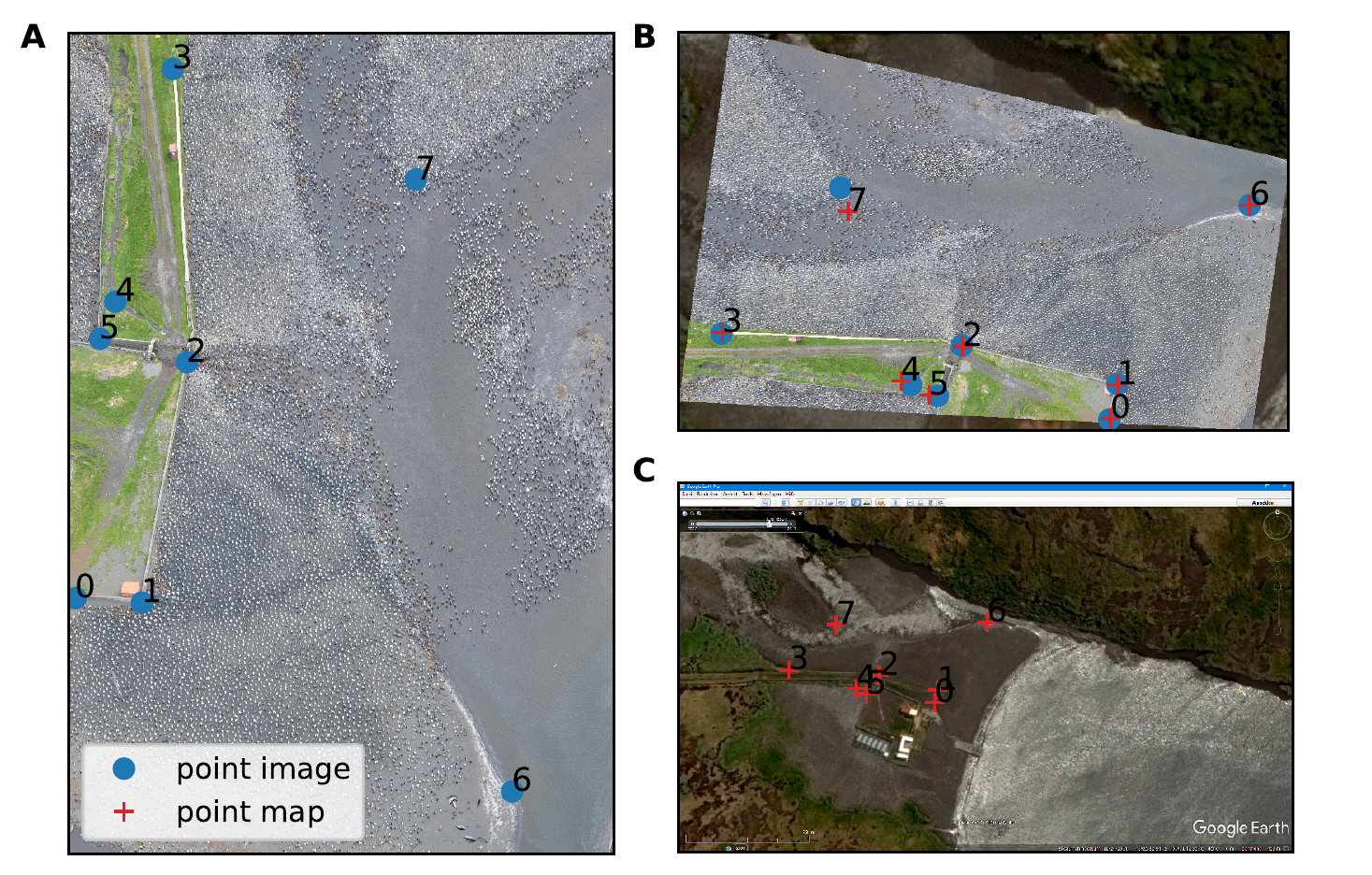}
		\caption{\textbf{Fit of image to map.}\newline A) recorded camera image from a helicopter flight at the Baie du Marin colony at the Crozet islands \cite{LeBohec2013}. B) image fitted over points in the image (blue) with points in the map image (red). C) a satellite image provided by Google Earth.}
		\label{fig:mapFit}
	\end{figure}
	
	For large tilt angles, e.g. if images are taken by a helicopter (Fig.~\ref{fig:mapFit}A), the size of the objects in the image does not vary sufficiently with the y position in the image so that the fitting approach based on the known object size is not viable.  In addition, the horizon unlikely to be visible. For such images, a different method is needed. If the approximate x,y location of the camera is known and an accurate map or a satellite image is available, point correspondences between the image and the map can be used to estimate the camera parameters using a process known as image registration. 
	
	In the example shown in Fig.~\ref{fig:mapFit} where we photograph a King penguin colony at Baie du Marin from a helicopter flying approximately 300 m above ground, we use eight points that are recognizable in the camera image and a satellite image provided by Google Earth  (Fig.~\ref{fig:mapFit}A,C). The cost function for our image registration is the distance between the projection of the image points to real-world coordinates  and the corresponding points in the satellite image. The fit routine then computes the \textit{height} and \textit{tilt} of the camera as well as the xy-position and heading angle. The example in Fig.~\ref{fig:mapFit} demonstrates that the fit routine matches all points except point \#7, which is the branch point of a river that likely has shifted from the time the satellite image was taken (Fig.~\ref{fig:mapFit}B).
	
	\section{Summary}
	We present a python package  for estimating extrinsic camera parameters based on image features, for image geo-referencing and correcting for perspective image distortions. The package is designed to assist in analysing images for ecological applications.  The package is published under the GPLv3 open source license to allow for continuous use and application in science. The documentation is hosted on \url{http://cameratransform.readthedocs.io} with explanations on how to install the package and with examples on how to use it. 
	
	\section{Acknowledgements}
	This work was supported by the Institut Polaire Français Paul-Emile Victor (IPEV, Programs no. 137 to CLB and 354 to FB). This study was funded by the Deutsche Forschungsgemeinschaft (DFG) grant FA336/5-1 and ZI1525/3-1 in the framework of the priority program "Antarctic research with comparative investigations in Arctic ice areas". 
	\bibliographystyle{plain}
	\bibliography{references}

\begin{thebibliography}{1}

\bibitem{10.2307/3784076}
Tricia~L Cutler and Don~E Swann.
\newblock {Using Remote Photography in Wildlife Ecology: A Review}.
\newblock {\em Wildlife Society Bulletin (1973-2006)}, 27(3):571--581, 1999.

\bibitem{Gregory2014}
Tremaine Gregory, Farah Carrasco~Rueda, Jessica Deichmann, Joseph Kolowski, and
  Alfonso Alonso.
\newblock {Arboreal camera trapping: taking a proven method to new heights}.
\newblock {\em Methods in Ecology and Evolution}, 5(5):443--451, 5 2014.

\bibitem{hartley2003multiple}
Richard Hartley and Andrew Zisserman.
\newblock {\em {Multiple view geometry in computer vision}}.
\newblock Cambridge University Press, Cambridge, 2003.

\bibitem{LeBohec2013}
Céline Le~Bohec.
\newblock {Programme 137 of the Institut Polaire Fran{\c{c}}ais Paul-Emile
  Victor (PI: C{\'{e}}line Le Bohec)}, 2013.

\bibitem{Lynch2015}
Tim~P. Lynch, Rachael Alderman, and Alistair~J. Hobday.
\newblock {A high-resolution panorama camera system for monitoring colony-wide
  seabird nesting behaviour}.
\newblock {\em Methods in Ecology and Evolution}, 6(5):491--499, 5 2015.

\bibitem{mobius1827barycentrische}
August~Ferdinand M{\"{o}}bius.
\newblock {\em {Der barycentrische Calcul, ein H{\"{u}}lfsmittel zur
  analytischen Behandlung der Geometrie}}.
\newblock Barth, Leipzig, 1827.

\bibitem{Zitterbart2011}
Daniel~P Zitterbart, Barbara Wienecke, James~P Butler, and Ben Fabry.
\newblock {Coordinated movements prevent jamming in an Emperor penguin huddle.}
\newblock {\em PloS one}, 6(6):e20260, 1 2011.

\end{thebibliography}
	
	\section*{Appendix}
	\subsection*{A. Backtransform for $x_3$=0}
	
	\begin{align}
		\begin{pmatrix}
			y_1 \\
			y_2 \\
			1
		\end{pmatrix}
		&=   \begin{pmatrix}
			c_{11} & c_{12} & c_{13} & c_{14}\\
			c_{21} & c_{22} & c_{23} & c_{24}\\
			c_{31} & c_{32} & c_{33} & c_{34}
		\end{pmatrix} \cdot
		\begin{pmatrix}
			s\cdot x_1 \\
			s\cdot x_2 \\
			s\cdot x_3\\
			s
		\end{pmatrix}\\
		& =   \begin{pmatrix}
			c_{11} & c_{12} & c_{13} \cdot x_3 & c_{14}\\
			c_{21} & c_{22} & c_{23} \cdot x_3 & c_{24}\\
			c_{31} & c_{32} & c_{33} \cdot x_3 & c_{34}
		\end{pmatrix} \cdot
		\begin{pmatrix}
			s\cdot x_1 \\
			s\cdot x_2 \\
			s\\
			s
		\end{pmatrix}\\
		& =   \begin{pmatrix}
			c_{11} & c_{12}  & c_{13} \cdot x_3+c_{14} \\
			c_{21} & c_{22}  & c_{23} \cdot x_3+c_{24} \\
			c_{31} & c_{32}  & c_{33} \cdot x_3+c_{34}
		\end{pmatrix} \cdot
		\begin{pmatrix}
			s\cdot x_1 \\
			s\cdot x_2 \\
			s
		\end{pmatrix}\\
		&= \tilde{C}  \begin{pmatrix}
			s\cdot x_1 \\
			s\cdot x_2 \\
			s
		\end{pmatrix}\\
		\tilde{C}^{-1} \cdot \begin{pmatrix}
			y_1 \\
			y_2 \\
			1
		\end{pmatrix}& =  \begin{pmatrix}
			s\cdot x_1 \\
			s \cdot x_2 \\
			s
		\end{pmatrix}
	\end{align}
\end{document}